\documentclass[twocolumn,prd,aps,showpacs,tightenlines,nofootinbib,preprintnumbers,
]{revtex4}
\usepackage{graphicx}
\usepackage{amssymb,latexsym}
\usepackage{amsfonts}
\usepackage{amssymb}
\usepackage{cancel}
\usepackage{color}
\newcommand{\lsim}{\mathrel{\hbox{\rlap{\lower.55ex\hbox{$\sim$}} \kern-.3em \raise.4ex \hbox{$<$}}}}
\newcommand{\gsim}{\mathrel{\hbox{\rlap{\lower.55ex\hbox{$\sim$}} \kern-.3em \raise.4ex \hbox{$>$}}}}
\newcommand{\bib}{\bibitem}
\newcommand{\beq}{\begin{equation}}
\newcommand{\beqa}{\begin{eqnarray}}
\newcommand{\eeq}{\end{equation}}
\newcommand{\eeqa}{\end{eqnarray}}

\newcommand{\ka}{\kappa}
\newcommand{\vp}{\varphi}
\newcommand{\Tcos}{T^\mathrm{cos}}
\newcommand{\Tpert}{T^\mathrm{s}}

\def\NATT#1#2#3{Nature (London) {\bf #1}, #2 (20#3)}

\def\frro{f_{RR0}}
\def\fro{f_{R0}}
\def\fo{f_0}

\begin{document}
\preprint{astro-ph/yymmnnn}

\title{Solar System constraints to general $f(R)$ gravity}

\author{Takeshi Chiba$^1$, Tristan L. Smith$^2$, and  Adrienne L. Erickcek$^2$ }%
\affiliation{
$^1$Department of Physics, College of Humanities and Sciences, \\
Nihon University, 
Tokyo 156-8550, Japan}
\affiliation{$^2$California Institute of Technology, Mail Code
130-33, Pasadena, CA 91125
}
\date{\today}


\begin{abstract}
It has been proposed that cosmic acceleration or inflation can be driven by replacing the Einstein-Hilbert action of general relativity with a function $f(R)$ of the Ricci scalar $R$.  Such $f(R)$ gravity theories have been shown to be equivalent to scalar-tensor theories of gravity that are incompatible with Solar System tests of general relativity, as long as the scalar field propagates over Solar System scales. Specifically, the PPN parameter in the equivalent scalar-tensor theory is $\gamma=1/2$, which is far outside the range allowed by observations.  In response to a flurry of papers that questioned the equivalence of $f(R)$ theory to scalar-tensor theories, it was recently shown explicitly, without resorting to the scalar-tensor equivalence, that the vacuum field equations for $1/R$ gravity around a spherically symmetric mass also yield $\gamma= 1/2$.  Here we generalize this analysis to $f(R)$ gravity and enumerate the conditions that, when satisfied by the function $f(R)$, lead to the prediction that 
$\gamma=1/2$.
\end{abstract}

\maketitle

\section{Introduction}

The evidence that the expansion of the Universe is currently accelerating \cite{Perlmutter:1998np,Riess:1998cb} suggests that the Universe is dominated by dark energy with a large negative pressure.  The predominant hypothesis is that a nonzero vacuum energy drives the acceleration, but this poses two serious theoretical questions: why is the vacuum energy nonzero, and why is it so miniscule?  An equally plausible alternative to dark energy is a modification of general relativity that would generate cosmic acceleration \cite{cdtt,Capozziello:2003tk}.  Modifying general relativity in this manner eliminates the need for dark energy, but it does not explain why the vacuum energy is zero.  Similar modifications of general relativity have also been proposed to drive inflation \cite{starobinsky}.

A possible modification to general relativity that generates an accelerated expansion is $1/R$  gravity \cite{cdtt}, in which a term proportional to $1/R$, where $R$ is the Ricci scalar, is added to the Einstein-Hilbert action so that the $1/R$ term dominates as the Hubble parameter decreases.  Soon after the introduction of this theory, it was shown that $1/R$ gravity is dynamically equivalent to a scalar-tensor gravity with no scalar kinetic term \cite{chiba}.   Moreover, the equivalence to scalar-tensor gravity applies to all modified gravity theories that replace the Einstein-Hilbert action with some function of the Ricci scalar [known as $f(R)$ gravity], provided that $f(R)$ has a nonzero second derivative with respect to $R$.  When the scalar field is light, this theory makes predictions that are incompatible with Solar System tests of general relativity \cite{Brans:1961sx, gamma, shapiro}.  Consequently, Ref.~\cite{chiba} concluded that a broad class of $f(R)$ gravity theories, including $1/R$ gravity, are ruled out by Solar System tests.  

Since then, however, the results in Ref.~\cite{chiba} were criticized by a number of papers \cite{pap1,pap2,pap3,pap4,pap5} and some even claim that Solar System experiments do not rule out any form of $f(R)$ gravity.  The essence of the criticism is that $f(R)$ gravity admits the Schwarzschild-de Sitter solution and hence the vacuum spacetime in the Solar System is not different from that in general relativity, although there were also broader objections to the equivalence between $f(R)$ and scalar-tensor gravity \cite{pap3}.   Working directly with the field equations, a recent paper \cite{smith} found that even though the Schwarzschild-de Sitter metric is a vacuum solution in $1/R$ gravity, it does not correspond to the solution around a spherically symmetric massive body.\footnote{Eddington made a similar mistake in $R^2$ gravity \cite{eddington}, which was later corrected by \cite{ps}.}   They found that the solution for the Solar System is identical to the spacetime derived using the corresponding scalar-tensor theory.

In this paper, we generalize the analysis of Ref.~\cite{smith} to a broad class of $f(R)$ gravities, namely those theories that admit a Taylor expansion of $f(R)$ around the background value of the Ricci scalar.  We work in the metric formalism, where the field equations are obtained by varying the action with respect to the metric and treating the Ricci scalar as a function of the metric.  The Palatini formalism, which treats the Ricci scalar as a function of the connection and varies the action with respect to the connection and the metric independently, yields different field equations for $f(R)$ gravity and has been studied extensively elsewhere (e.g. Refs.~\cite{pal1,pal2,pal3,pal4,pal5}).

This paper is organized as follows: In Section \ref{sec:spherical}, we solve the linearized field equations around a spherical mass and find that the solution in the Solar System is in agreement with the solution obtained using the equivalent scalar-tensor theory.  When $f(R)$ satisfies a condition that is analogous to the scalar field being light in the equivalent scalar-tensor theory, the resulting spacetime is incompatible with Solar System tests of general relativity.  In Section \ref{sec:cases}, we consider how our analysis applies to several $f(R)$ gravity theories, including general relativity.  This particular example illustrates the connection between $f(R)$ gravity and general relativity and clarifies the requirements for a general relativistic limit of an $f(R)$ theory.   We summarize our conclusions in Section \ref{sec:conclusions}.

\section{Weak-Field Solution around a Spherical Star \label{sec:spherical}}
We consider gravitational theories with actions of the form
\beq
S=\frac{1}{2\kappa} \int d^4x \sqrt{-g}f(R)+S_m,
\label{action}
\eeq
where $f(R)$ is a function of the Ricci scalar $R$ and $S_m$ is the matter action. 
The field equation obtained by varying the action with respect to the metric is
\beqa
f_RR_{\mu \nu}-{1\over 2}f g_{\mu \nu}-\nabla _{\mu}\nabla _{\nu} f_R+\Box f_Rg_{\mu \nu}=\kappa T_{\mu \nu},
\label{eom}
\eeqa
where $f_R \equiv df/dR$. 
In previous studies, predictions of Solar System dynamics in these theories were analyzed 
by appealing to an equivalence with scalar-tensor theories \cite{chiba}.  We review this equivalence
in Appendix \ref{appendix}.  Since the equivalent scalar-tensor theory is incompatible with Solar System observations if the scalar field propagates on Solar System scales, Ref.~\cite{chiba} concluded that the corresponding $f(R)$ theories are ruled out.  We now show that this conclusion can be made \emph{without} appealing to the equivalence between $f(R)$ and scalar-tensor gravity.  Instead, we work directly with the linearized field equations about a spherical mass distribution.  Our treatment clarifies and amends a similar analysis presented in Ref. \cite{Jin:2006if}, and we extend it to cases where the background value of the Ricci scalar equals zero. 

We now find the metric that describes the spacetime around a spherical body in $f(R)$ gravity in the weak-field regime.   To do this, we must choose a background spacetime around which to linearize the field equations.  The only physically relevant choice is an isotropic and homogeneous background spacetime that solves Eq.~(\ref{eom}) for some spatially uniform cosmological stress-energy tensor $\Tcos_{\mu\nu}$.  The evolution of the time-dependent and spatially homogeneous background scalar curvature $R_0(t)$ is determined by the trace of Eq.~(\ref{eom}), 
\begin{equation}
\fro(t) R_0(t)-2 \fo(t)+ 3 \Box \fro(t)=\kappa \Tcos(t),
\label{back}
\end{equation}
where $\fro \equiv df/dR|_{R=R_0}$, $f_0 \equiv f(R_0)$ and $\Tcos \equiv g^{\mu\nu}\Tcos_{\mu\nu}$.  

In order to investigate perturbations away from this background, we express the Ricci scalar as the sum of two components: 
\beq
R(r,t) \equiv R_0(t)+R_1(r),
\eeq
where $R_0(t)$ is the spatially homogenous background curvature that solves Eq.~(\ref{back}) and $R_1(r)$ is a time-independent perturbation to this background curvature.  We assume that all derivatives of $f(R)$ are well-defined at the present-day value of $R_0$ so that we may use a Taylor expansion of $f(R)$ around $R=R_0$ to evaluate $f(R_0+R_1)$ and $f_{R}(R_0+R_1)$.   We will terminate the expansion by neglecting terms nonlinear in $R_1$.   Provided that the higher-order terms of the Taylor series do not cancel in some contrived way, neglecting the higher-order terms is only justified if the sum of the zeroth-order and linear terms is greater than all other terms in the Taylor expansion.  Specifically, we require that
\begin{eqnarray}
\fo + \fro R_1 & \gg& \frac{1}{n!} f^{(n)}(R_0) R_1^n, \label{taylorcond1}\\
\fro + \frro  R_1 &\gg& \frac{1}{n!} f^{(n+1)}(R_0) R_1^n, \mbox{   for all } n>1, \label{taylorcond2}
\end{eqnarray}
where $\frro \equiv d^2f/dR^2|_{R=R_0}$ and $f^{(n)}(R_0) \equiv d^nf/dR^n|_{R=R_0}$.  

Now we consider the trace of Eq. (\ref{eom}) with both a cosmological matter source described by $\Tcos$ and a finite, time-independent, spherically symmetric matter source, described by $\Tpert$:
\begin{equation}
f_R R-2 f+ 3 \Box f_R=\kappa \left(\Tcos+\Tpert \right).
\label{traceeq}
\end{equation}
Using first-order Taylor expansions to evaluate $f_R$ and $f$ and neglecting ${\cal O}(R_1^2)$ terms, we obtain a linearized version of Eq.~(\ref{traceeq}):
\begin{eqnarray}
&& 3 \frro \Box R_1(r) - \bigg[\fro(t)  \nonumber \\ 
&&- \frro(t) R_0(t)-3 \Box \frro(t)\bigg] R_1 =
\kappa \Tpert.
\label{trace}
\end{eqnarray}
To obtain this equation, we used the fact that $R_0(t)$ solves Eq.~(\ref{back}) to eliminate terms that are independent of $R_1$.   
By dropping ${\cal O}(\frro R_1^2)$ terms from Eq.~(\ref{trace}) while keeping the $\frro R_0 R_1$ term, we have implicitly assumed that $R_1 \ll R_0$ if $R_0$ is nonzero.  We will check that this condition is satisfied after the discussion following Eq.~(\ref{phi}).  If $R_0$ is zero, then the ${\cal O}(\frro R_1^2)$ is guaranteed to be smaller than the nonzero terms in Eq.~(\ref{trace}) by virtue of Eq.~(\ref{taylorcond1}).
Note that if $f_{RR0}=0$, as in general relativity, this equation becomes simply $f_{R0}R_1= -\kappa \Tpert$.   If in addition $f_{R0}$ is nonzero then $R_1$ must vanish outside the star and hence the Schwarzschild-de Sitter solution becomes the solution to the field equation outside the source. However, if $\frro \neq 0$, this is no longer necessarily the case.  
    
Finally, we take our background metric to be a flat Friedmann-Robertson-Walker (FRW) metric.  We then consider a spherically symmetric perturbation to this background so that the linearized perturbed metric takes the form
\begin{eqnarray}
ds^2&=&-[1+2\Psi(r)]dt^2 \nonumber \\
&&+a(t)^2\big\{[1+2\Phi(r)]dr^2+r^2d\Omega^2\big\},
\label{metric}
\end{eqnarray}
where the present value of $a(t)$ is one. When solving the field equations, we will keep only terms linear in the perturbations $\Psi$ and $\Phi$.   


We will now solve Eq.~(\ref{trace}) for a nonzero $\frro$.  Since we confine our analysis to a static perturbation $R_1(r)$, $\Box$ becomes the flat-space Laplacian operator $\nabla^2$.  Restricting our analysis to a source with mass density $\rho(r)$ and negligible pressure, we may rewrite Eq.~(\ref{trace}) as
\begin{equation}
\nabla^2 R_1 - m^2 R_1 = - \frac{\kappa \rho}{3 \frro},
\label{diffeq}
\end{equation}
where we have defined a mass parameter
\begin{equation}
m^2 \equiv \frac{1}{3} \left(\frac{\fro}{\frro} - R_0 - 3 \frac{ \Box \frro}{\frro} \right).
\label{effmass}
\end{equation}
Due to the evolution of $R_0(t)$, this mass parameter varies in time.  However, the time-scale of variation in the cosmological background spacetime is comparable to the current Hubble time.  Since this time-scale is much longer than the time-scale of Solar System dynamics, we may neglect the time variation of the background spacetime when considering the behavior of bodies within the Solar System \cite{will}.  Therefore, for the purposes of this calculation, we take $m$ to be time-independent. 

The Green's function $G(r)$ for this differential equation depends on the sign of $m^2$:
\begin{equation}
G(r) =  \left\{ \begin{array}{ll}
-\cos(mr)/(4 \pi r) & \ m^2 < 0,\\
-\exp(-mr)/(4 \pi r) & \ m^2 > 0,\\
\end{array}
\right.
\end{equation}
where $m \equiv \sqrt{|m^2|}$.  If $mr \ll 1$, then both Green's functions are approximately $-1/(4\pi r)$, which is the Green's function for Laplace's equation.  In this case, the term proportional to $m^2$ in Eq.~(\ref{diffeq}) may be neglected and the solution outside the star is given by
\begin{equation}
R_1 = \frac{\kappa}{12\pi \frro}\frac{M}{r},
\label{sol}
\end{equation}
where $M$ is the total mass of the source. 
We note that when applied to $1/R$ gravity with a static de Sitter background, this result agrees with the result presented in Ref.~\cite{smith}.  

We emphasize that in order for this solution for $R_1$ to be valid, we must have $mr \ll 1$.   Only when this condition is satisfied is the trace of the field equation well-approximated by Laplace's equation.  This restriction was not mentioned in Ref.~\cite{Jin:2006if}.  The physical interpretation of this constraint is clear when one considers the equivalent scalar-tensor theory.  When one switches to a frame where the scalar degree of freedom is canonical, the effective mass of the scalar field evaluated in the Jordan frame is \cite{chiba}
\begin{equation}
m_\varphi^2 = \frac{\fro}{3} \left( \frac{1}{\frro} + \frac{R_0}{\fro} - \frac{4\fo}{(\fro)^2} -\frac{2 \kappa T^{\mathrm{cos}}}{(\fro)^2}\right).
\label{mphi}
\end{equation}
Since $R_0$ is the solution to Eq.~(\ref{back}), this expression may be simplified to
\begin{eqnarray}
m_\varphi^2 = \frac{1}{3} \left({\frac{\fro}{\frro} - R_0- 6\frac{\Box \fro}{\fro}}\right) \label{scalar_mass}.
\label{mass}
\end{eqnarray}
It is clear that both $m_{\varphi}$ and $m$ [defined by Eq.~(\ref{effmass})] are of the same order.  Therefore, the condition that $mr \ll 1$ is equivalent to demanding that the scalar field be light ($m_\varphi r \ll 1$).   See Appendix A for more details. 

In summary, Eq.~(\ref{sol}) is a solution to the trace of the field equation within the Solar System only if \emph{the scalar degree of freedom propagates on Solar System scales}.  In terms of $f(R)$, the necessary condition is
\begin{equation}
|m^2|r^2 \equiv \left|{\frac{1}{3} \left({\frac{\fro}{\frro} - R_0-3\frac{\Box \frro}{\frro}}\right)}\right|r^2 \ll 1.
\label{tracecond}
\end{equation}
The triangle inequality tells us that the mass constraint given by Eq.~(\ref{tracecond}) implies that
\begin{equation}
\left|\frac{\fro}{\frro}\right|r^2 - \left| R_0 - 3\frac{\Box \frro}{\frro} \right| r^2  \ll 1.
\end{equation}
Finally, since $\Box \frro/\frro \sim H^2$, where $H \equiv \dot{a}/a$ is the current Hubble parameter, and we know that $R_0 r^2 \sim H^2 r  \ll 1$ by cosmological constraints, the mass constraint implies that
\begin{equation}
\left|\frac{\fro}{\frro}\right|r^2 \ll 1.
\label{masscond}
\end{equation}


We will now use the expression for $R_1$ given by Eq.~(\ref{sol}) to solve the field equations for the metric perturbations $\Psi$ and $\Phi$.  As we did for the trace of the field equation, we simplify the field equations by replacing $f(R)$ and $f_R (R)$ with first-order Taylor expansions around the background value $R_0$ to obtain field equations that are linear in $R_1$.  Using Eq.~(\ref{back}) to simplify this expression, we obtain
\begin{eqnarray}
\fro(R^\mu_\nu - [R_0]^{\mu}_{\nu}) + \frro R_1 R^\mu_\nu - \frac{1}{2} \fro R_1 \delta^\mu_\nu  && \label{linfield}
\\
- \frro \nabla^\mu \nabla_\nu R_1 + \delta^\mu_\nu \frro \Box R_1 &=&  \kappa {\Tpert}^{\mu}_{\nu}, \nonumber
\end{eqnarray}
where $[R_0]^{\mu}_{\nu}$ is the unperturbed FRW Ricci tensor and $\delta^\mu_\nu$ is the Kronecker delta.  We neglected time derivatives of the background metric when deriving this equation.  As previously noted, the time-scale of variations in $R_0$ is much longer than that of Solar System dynamics, making the terms involving time derivatives of $R_0$ irrelevant to gravitational effects within the Solar System.

We simplify Eq.~(\ref{linfield}) further by dropping several negligible terms.  We continue to ignore terms that depend on the variation of the background spacetime by dropping terms that involve products of $\Phi$, $\Psi$ and $\frro R_1$ with $H$ and $dH/dt$.
Since we are working in the weak-field regime, we neglect all terms that are nonlinear functions of the metric perturbations $\Phi$ and $\Psi$.   Keeping only terms that are linear in $\Phi$ and $\Psi$ allows us to replace the $\Box$ with the flat-space Laplacian operator $\nabla^2$ since the perturbation is assumed to be static.
Finally, we know from Eq.~(\ref{sol}) that $\frro R_1 \sim \kappa M/r$, and we expect $\Psi$ and $\Phi$ to be proportional to $\kappa M/r$ as well.  Therefore, $\frro R_1 \Psi$ and $\frro R_1 \Phi$ are second-order quantities, and we may neglect them.  With these simplifications, the $tt,rr,\theta\theta$ components of Eq.~(\ref{linfield}) are respectively
\beqa
\fro \nabla^2\Psi+\frac{1}{2}f_{R0}R_1-f_{RR0}\nabla^2R_1&=&\kappa{\rho}, \label{eom1}\\
\fro \left(-\Psi''+\frac{2}{r}\Phi'\right)-\frac{1}{2}\fro R_1+
\frac{2}{r}\frro R_1^\prime&=&0,  \label{eom2}\\
f_{R0}\left({1\over r}\Phi'-{1\over r}\Psi'+{2\over r^2}\Phi\right)&& \nonumber\\
-{1\over 2}f_{R0}R_1
+{1\over r}f_{RR0}R_1^\prime+f_{RR0}R_1^{\prime\prime}&=&0,\label{eom3}
\eeqa
where the prime denotes differentiation with respect to $r$.  The $\phi\phi$ component of Eq.~(\ref{linfield}) is identical to the $\theta\theta$ component given by Eq.~(\ref{eom3}).

Recalling that $R_1$ solves Eq.~(\ref{diffeq}) with $m^2 = 0$ so that $\nabla^2 R_1$ is proportional to the density $\rho$,  Eq.~(\ref{eom1}) may be rewritten 
\beq
f_{R0}\nabla^2\Psi=\frac{2}{3} \kappa \rho - \frac{1}{2}f_{R0}R_1.
\eeq
We express $\Psi$ as the sum of two functions: $\Psi = \Psi_0 + \Psi_1$, where
\beqa
f_{R0}\nabla^2\Psi_0&=&\frac{2}{3} \kappa \rho, \label{psi0eqn}\\
f_{R0}\nabla^2\Psi_1 &=&  - \frac{1}{2}f_{R0}R_1. \label{psi1eqn}
\eeqa
Provided that $\fro \neq 0$, Eq.~(\ref{psi0eqn}) may be integrated via Gauss's Law to give
\beq
\Psi_0^\prime (r) = \frac{\kappa}{6\pi \fro} \frac{m(r)}{r^2},
\label{psiprime}
\eeq
where $m(r)$ is the mass enclosed in a sphere of radius $r$.  If we assume that $\Psi_0$ vanishes as $r \rightarrow \infty$, we may integrate Eq.~(\ref{psiprime}) to obtain
\beq
\Psi_0=-\frac{\kappa}{6\pi \fro}  {M\over r},
\label{psi}
\eeq
outside the star.  Solving Eq.~(\ref{psi1eqn}) outside the star using Eq.~(\ref{sol}) for $R_1$ yields 
\beq
|\Psi_1| = \frac{1}{48\pi \frro} \kappa Mr  \ll \frac{1}{\fro} \frac{\kappa M}{r},
\eeq
where the inequality follows from Eq.~(\ref{masscond}).  Since $\Psi_0 \sim \kappa M/(\fro r)$ outside the star we have shown that $|\Psi_1| \ll |\Psi_0|$.  Therefore, we may neglect $\Psi_1$ and conclude that $\Psi=\Psi_0$ as given by Eq.~(\ref{psi}).  This expression for $\Psi$ is used to define Newton's constant: $G \equiv \kappa/ (6 \pi f_{R0})$.   For $1/R$ gravity with a static vacuum de Sitter background, $\fro = 4/3$, so $\kappa$ takes its standard value of $8\pi G$ and Eq.~(\ref{psi}) matches the corresponding result in Ref.~\cite{smith}.

We now turn our attention to Eq.~(\ref{eom2}), which we will solve for $\Phi$.  First, we note that Eq.~(\ref{sol}) implies that $R_1^\prime = -R_1/r$.  Therefore, the ratio of the second two terms in Eq.~(\ref{eom2}) is 
\beq
\left|\frac{(1/2) \fro R_1}{2\frro R_1^\prime/r}\right| \sim \left|\frac{\fro}{\frro}\right| r^2 \ll 1,
\eeq
where the inequality follows from Eq.~(\ref{masscond}).  Consequently, the $\fro R_1$ term is negligible, and we drop it from the equation.  Differentiating Eq.~(\ref{psiprime}) to find $\Psi^{\prime\prime}$, and using Gauss's Law to obtain $R_1^{\prime}$ from Eq.~(\ref{diffeq})  (with $m^2 = 0$), we may then rewrite Eq.~(\ref{eom2}) as
\beq
\Phi^\prime(r) = \frac{\kappa}{12\pi \fro} \frac{\rm d}{{\rm d}r} \left(\frac{m(r)}{r}\right).
\label{phiprime}
\eeq
Assuming that $\Phi$ vanishes as $r \rightarrow \infty$, this equation may be integrated to obtain
\beq
\Phi=\frac{\kappa}{12\pi f_{R0}} \frac{M}{r},
\label{phi}
\eeq
outside the star.  It is easy to verify that Eqs. (\ref{psi}) and (\ref{phi}) also satisfy the third field equation, Eq.~(\ref{eom3}). 

We may now check our assumption that $R_1 \ll R_0$ for nonzero $R_0$.  From the expression for $R_1$ given by Eq.~(\ref{sol}) and our definition that $\kappa \equiv 6 \pi \fro G$, we see that 
\begin{equation}
\frac{R_1}{R_0} \lsim \frac{1}{R_0}  \left(\frac{GM}{R_\mathrm{s}}\right)\frac{\fro}{\frro},
\end{equation}
where $R_\mathrm{s}$ is the radius of the star.  It is easy to check that this expression holds inside the star as well by integrating Eq.~(\ref{diffeq}) into the interior of the star.   Therefore, our assumption that $R_1 \ll R_0$ places an additional condition on the ratio $\fro/\frro$:
\begin{equation}
\left| \frac{\fro}{\frro} \right| \ll R_0 \left(\frac{R_\mathrm{s}}{GM}\right) \mbox{ for } R_0 \neq 0.
\label{R1cond}
\end{equation}
If $\fro/\frro \sim R_0$, as is the case for many $f(R)$ theories with nonzero $R_0$, then this condition is always satisfied.

Thus we have shown explicitly that $\Psi=-2\Phi=-GM/r$ for all $f(R)$ theories with nonzero $\frro$ that satisfy the conditions given by Eqs.~(\ref{taylorcond1}), (\ref{taylorcond2}), (\ref{tracecond}) and (\ref{R1cond}).  Transforming the metric given by Eq.~(\ref{metric}) to isotropic coordinates, taking $a = 1$ today, and keeping only terms that are linear in $GM/r$ gives
\begin{eqnarray}
ds^2=&&-\left(1-\frac{2GM}{r}\right)dt^2\nonumber\\
&&+\left(1+\frac{GM}{r}\right)\left[dr^2+r^2d\Omega^2\right].
\label{isometric}
\end{eqnarray}
It is clear that this spacetime is equivalent to a Parameterized Post-Newtonian spacetime with PPN parameter $\gamma = 1/2$.  This result is in gross violation of observations; Solar System tests require that $\gamma=1+(2.1\pm 2.3)\times 10^{-5}$ \cite{gamma,shapiro}.  We also note that this result is in precise agreement with the results 
obtained using the equivalent scalar-tensor theory \cite{chiba} (see also \cite{olmo}). 

\section{Case Studies \label{sec:cases}}

First, we show how we regain the results of general relativity if we take $\frro = 0$ and assume that our linearized Taylor expansion is a valid approximation.  We note that general relativity [$f(R) = R$] satisfies both of these conditions.  

Taking $\frro=0$, Eq.~(\ref{trace}) yields 
\beq
f_{R0}R_1= \kappa \rho.
\label{r12}
\eeq 
When $\frro=0$, the $\fro R_1$ terms in the field equations [Eqs.~(\ref{eom2}-\ref{eom3})] are no longer negligible compared to the terms proportional to $\frro$ since these terms vanish.  The field equations then become
\beqa
f_{R0}\nabla^2\Psi+{1\over 2}f_{R0}R_1&=&\kappa \rho,\label{eom12}\\
f_{R0}\left(-\Psi''+{2\over r}\Phi'\right)-{1\over 2}f_{R0}R_1&=&0,\label{eom22}\\
f_{R0}\left({1\over r}\Phi'-{1\over r}\Psi'+{2\over r^2}\Phi\right)
-{1\over 2}f_{R0}R_1&=&0.\label{eom32}
\eeqa
Using Eq.~(\ref{r12}), Eq.~(\ref{eom12}) becomes
\beq
f_{R0}\nabla^2\Psi=\frac{\kappa}{2}{\rho},
\eeq
and the solution outside the star is
\beq
\Psi=-{\kappa \over 8\pi f_{R0}}{M\over r}.
\eeq
From Eq.~(\ref{eom22}) and Eq.~(\ref{eom32}), we have
\beq
{f_{R0}\over r^2}\left(r\Phi\right)'=\frac{\kappa}{2} \rho,
\eeq
and the solution outside the star is 
\beq
\Phi={\kappa\over 8\pi f_{R0}}{M\over r}=-\Psi.
\eeq
Since $\Psi = -\Phi = -GM/r$, transforming to isotropic coordinates reveals that $\gamma = 1$ as expected.

With this result it is easy to see why the $\mu \rightarrow 0$ limit in $1/R^n\ (n>0)$ gravity \emph{does not} recover general relativity.  In $1/R^n$ gravity \cite{cdtt}, we have
\begin{equation}
f(R) = R- \frac{\mu^{2+2n}}{R^n}, \ \ n>0.
\label{gen_1_o_R}
\end{equation} 
The static solution to Eq.~(\ref{back}) with $\Tcos=0$ is $R_0 = (n+2)^{1/(n+1)} \mu^2$, and $\frro \propto \mu^{-2}$.  Therefore, $\frro$ \emph{diverges rather than vanishes} in the limit that $\mu\rightarrow 0$, and general relativity is \emph{not} regained.   The mass parameter for this theory has the dependence $m^2 \propto \mu^{2}$ and hence it  vanishes in the limit that $\mu\rightarrow 0$.  Furthermore, a Taylor series of Eq.~(\ref{gen_1_o_R}) around $R_0$ is well-behaved and cosmological constraints tell us that $\mu \sim H$ so that $m^2 r^2 \ll 1$ in the Solar System.  We conclude that the analysis of general $f(R)$ gravity given in Section \ref{sec:spherical} applies and $\gamma=1/2$ for these theories in a static background.

We note however that the static solution to Eq.~(\ref{back}) may not describe the current cosmological background in $1/R^n$ gravity.  This solution is unstable, and without fine-tuning of the initial conditions, this spacetime will evolve toward a spacetime with $R_0 \ll \mu^2$ \cite{cdtt}.  In that case, we note that
\begin{eqnarray}
\frac{(m!)^{-1}f^{(m)}(R_0)R^m_1}{\fo +\fro R_1} &\lsim& \left(\frac{GM}{r}\right)^m \ll 1,\\
\frac{(m!)^{-1}f^{(m+1)}(R_0)R^{m}_1}{\fro +\frro R_1} &\lsim& \left(\frac{GM}{r}\right)^m \ll 1,
\end{eqnarray}
so that Eqs.~(\ref{taylorcond1}) and (\ref{taylorcond2}) are still satisfied.  Furthermore, $m^2  \propto R_0$, so, as in the static-background case, the mass is of order the Hubble parameter today.  Therefore, the $\gamma = 1/2$ result holds even during the late-time evolution of $1/R^n$ gravity.

Next we consider Starobinsky gravity \cite{starobinsky} which has
\begin{equation}
f(R) = R+\frac{R^2}{\alpha^2}.
\end{equation}
The static solution to Eq.~(\ref{back}) with $\Tcos=0$ is $R_0 = 0$ for this theory.  Since $f(R)$ is a second-order polynomial, the first-order Taylor expansion of $f_R(R_0 + R_1)$ is exact.  The ${\cal{O}}(R_1^2)$ term in the Taylor expansion of $f(R_0 + R_1)$ is suppressed compared to the linear term by a factor of $GM/r$ and is therefore negligible.  The mass parameter for this theory is proportional to $\alpha^2$, so Eq.~(\ref{sol}) is a solution for $R_1$ if $\alpha^2 r^2 \ll 1$.  Therefore, $\gamma = 1/2$ in this theory if $\alpha^2 r^2 \ll 1$ inside the Solar System.  If the mass parameter $\alpha$ is made large (i.e.~if $\alpha \simeq 10^{12} $ GeV as proposed in Ref.~\cite{starobinsky}), then this condition is not satisfied and we cannot use the analysis in Section \ref{sec:spherical} to calculate $\gamma$ for this theory.

Next we consider an example of a theory that uses two mass parameters: a hybrid between Starobinsky gravity and $1/R$ gravity.  In particular, consider the function
\begin{equation}
f(R) = R + \frac{1}{\alpha^2} R^2 - \frac{\mu^4}{R}.
\end{equation}
We then find that, as in the usual $1/R$ case, we have $R_0 = \sqrt{3} \mu^2$ (for a static background in vacuum).  However, 
\begin{equation}
m^2 = 3 \mu^2 \left(\frac{\alpha^2}{9 \mu^2 - \sqrt{3} \alpha^2}\right).
\end{equation}
We can make this quantity as large as we want by letting the denominator tend towards zero, which gives the condition $\alpha \rightarrow 3^{3/4} \mu$.  Thus, in this model we can violate the conditions listed in Section \ref{sec:spherical} by fine-tuning the parameters.

Finally, we consider power-law gravitational actions \cite{cb}:
\begin{equation}
f(R) = \left(\frac{R}{\alpha}\right)^{1+\delta}.
\end{equation}
Assuming that $\delta \neq 1$, the static vacuum solution to Eq.~(\ref{back}) is $R_0 = 0$.  If $\delta$ is not an integer, there will be some derivative that is not defined at $R =0$, which causes the Taylor expansion to fail around that point.  In particular, if it is supposed that $\delta \ll 1$, then at least the second derivative will be undefined so that the Taylor expansion will fail.   For $\delta = 1$ the static vacuum background value $R_0$ is undetermined.  However, if we choose $R_0\neq 0$ then all of the conditions listed in Section \ref{sec:spherical} are satisfied and we conclude that $\gamma = 1/2$ in agreement with Ref.~\cite{bicknell}.  If $\delta$ is an integer greater than one, then the Taylor expansion around $f(R_0 = 0)$ is well-defined, but we cannot drop the terms that are nonlinear in $R_1$ since the linearized function vanishes.   Therefore, this analysis is incapable of determining whether $f(R) = R^{1+\delta}$ gravity with $\delta \neq 1$ conflicts with Solar System tests.

\section{Conclusions}
\label{sec:conclusions}

By analyzing the field equations around a spherically symmetric mass, we have shown that, in agreement with the analysis in Ref.~\cite{chiba}, the PPN parameter $\gamma$ of general $f(R)$ gravity is $\gamma=1/2$ given the following conditions:
\\
\\
I. The Taylor expansions of $f(R)$ and $df / dR$ about the current background value $R=R_0$, where $R_0$ solves Eq.~(\ref{back}), are well-defined and dominated by terms that are linear in deviations away from $R=R_0$.  If $R_0$ is non-zero, then the deviations from $R_0$ are small compared to $R_0$.  This condition may be re-expressed as Eq.~(\ref{R1cond}) and is closely related to the third condition stated below.
\\
\\
II. The second derivative of $f(R)$ with respect to $R$ is nonzero when evaluated at the background value of $R = R_0$.
\\
\\
III. The mass parameter given by Eq.~(\ref{effmass}) respects the condition $mr \ll 1$ within the Solar System.  
\\
\\
For theories with one extra mass parameter and non-zero $R_0$, as in $1/R$ gravity, it is reasonable to assume that $\fro/\frro \sim R_0$.  In that case, the latter part of the first condition is always true and the third condition is satisfied provided that $R_0 r^2 \ll 1$ within the Solar System.  However, for theories with multiple mass parameters, such as the Starobinsky-$1/R$ hybrid presented in this paper, it is possible that this condition can be violated. 

The second and third conditions listed above correspond to synonymous conditions in the scalar-tensor treatment: $f(R)$ and scalar-tensor gravity are equivalent only if the second derivative of $f(R)$ is nonzero, and $\gamma=1/2$ only if the scalar field is light enough to propagate through the Solar System.  Therefore, we have also verified that, contrary to the claim of some authors \cite{pap1,pap2,pap3,pap4,pap5}, calculating the Solar System predictions of $f(R)$ gravity using the equivalent scalar-tensor theory is a valid technique.

\begin{acknowledgments}
During the final preparations of this paper we learned of recent work along similar lines \cite{paplast}.
TLS and ALE thank Marc Kamionkowski for useful suggestions and conversations.  TC was supported in part by a Grant-in-Aid for Scientific 
Research (No.17204018) from the Japan Society for the Promotion of Science and in part by Nihon University.  TLS is supported in part by DoE DE-FG03-92-ER40701, NASA NNG05GF69G, and the Gordon and Betty Moore Foundation.  ALE acknowledges the support of an NSF Graduate Fellowship.  
\end{acknowledgments}

\appendix
\section{Review of scalar tensor equivalence \label{appendix}}

The action for the scalar-tensor theory that is equivalent to $f(R)$ gravity is
\beqa
S=\frac{1}{2\kappa}\int d^4x\sqrt{-g}\left[f(\phi)+f_{\phi}(\phi)(R-\phi)\right] + S_{m},
\label{action2}
\eeqa
where $f_{\phi}(\phi) \equiv df/d\phi$ and $S_m$ is the matter action. The field equation for $\phi$ is $\phi =R$ if $d^2f/d\phi^2 \neq 0$. Since the relation between $\phi$ and $R$ is 
purely algebraic, it can be resubstituted into the action to reproduce
the action for $f(R)$ gravity given by Eq.~(\ref{action}). 
After the conformal transformation
$g_{\mu\nu}^E \equiv f_{\phi}(\phi)g_{\mu\nu}$, 
the action becomes that of general relativity with a minimally coupled
scalar field:
\begin{widetext}
\begin{eqnarray}
S={1\over 2\ka}\int d^4x\sqrt{-g_E}\left(R_E-
{3\over 2f_{\phi}(\phi)^2}g_E^{\mu\nu}
[\nabla_{E\mu}f_{\phi}(\phi)][\nabla_{E\nu}f_{\phi}(\phi)]-{1\over f_{\phi}(\phi)^2}
\left[\phi f_{\phi}(\phi)-f(\phi)\right]
\right) +S_{m}.
\label{action3}
\end{eqnarray}
\end{widetext}
Introducing a canonical scalar field $\vp$ such that 
$f_{\phi}(\phi)=\exp(\sqrt{2\kappa/3}\vp)$, Eq.~(\ref{action3}) can be rewritten as 
\begin{equation}
S=\int d^4x\sqrt{-g_E}\left({1\over 2\ka}R_E-{1\over 2}(\nabla_E\vp)^2
-V(\vp)\right)\nonumber + S_{m},
\end{equation}
where the potential is defined by
\begin{equation}
V(\vp)\equiv\frac{\phi(\vp) f_{\phi}[\phi(\vp)]-f[\phi(\vp)]}
{2\ka f_{\phi}[\phi(\vp)]^2}.
\label{Vdef}
\end{equation}
The absence of the kinetic term in Eq.~(\ref{action2}) implies the Brans-Dicke parameter 
of $f(R)$ gravity theories is $\omega=0$ \cite{Brans:1961sx}.  From an analysis of 
Brans-Dicke gravity, if the scalar degree of freedom can propagate on scales much larger than 
the Solar System, we can 
conclude that $\gamma=(1+\omega)/(2+\omega)=1/2$ \cite{Brans:1961sx}. 

In the frame where $\varphi$ is canonical (the Einstein frame) $\varphi$ has the equation of motion
\begin{equation}
\Box_E \varphi = \frac{d V}{d\varphi} + \sqrt{\frac{\kappa}{6}} f^{\prime}(\phi)^{-2} T^{\mathrm{M}},
\label{einsteinframe}
\end{equation}
where the prime denotes differentiation with respect to $\phi$. 
When we re-express Eq.~(\ref{einsteinframe}) in terms of $f^{\prime}(\phi)$ and the usual metric $g_{\mu \nu}$, we recover Eq.~(\ref{traceeq}).  Therefore, we stress that this reformulation contains \emph{no new dynamics} compared to the expressions used in this paper.  The two formulations are entirely equivalent. 

In order to derive the mass $m_{\varphi}$, we let $\varphi = \varphi_0(t) + \varphi_1(r)$ and $T^{\mathrm{M}} = T^{\mathrm{cos}} + \Tpert$ so that $\varphi_0(t)$ satisfies Eq.~(\ref{einsteinframe}) with $T^{\mathrm{cos}}$.  We then expand to linear order in the perturbation $\varphi_1$, writing Eq.~(\ref{einsteinframe}) in terms of the physical metric $g_{\mu \nu}$.   We find
\begin{equation}
\Box \varphi_1 = f^{\prime}(\phi_0)\left(\frac{d^2 V}{d\varphi^2}\bigg|_{\varphi_0} - \frac{2}{3}\kappa \frac{T^{\mathrm{cos}}}{[f^{\prime}(\phi_0)]^2}\right)\varphi_1 +  \sqrt{\frac{\kappa}{6}} \frac{\Tpert}{f^{\prime}(\phi_0)},
\end{equation}
where $\phi_0$ denotes the background field value for the $\phi$ field.  Using Eq.~(\ref{Vdef}) to evaluate $d^2 V/d\varphi^2$, we have
\begin{equation}
m_\varphi^2 = \frac{f^{\prime}(\phi_0)}{3} \left[\frac{1}{f^{\prime \prime}(\phi_0)} + \frac{\phi_0}{f^{\prime}(\phi_0)} - \frac{4f(\phi_0)}{[f^{\prime}(\phi_0)]^2} - 2 \kappa \frac{T^{\mathrm{cos}}}{[f^{\prime}(\phi_0)]^2}\right].
\end{equation}
Finally, we may rewrite $m_\varphi^2$ as Eq.~(\ref{mphi}) since $\phi_0 = R_0$. 
We conclude that if $m_{\vp}^2 r^2 \ll 1$ then $\gamma = 1/2$ as discussed in Ref.~\cite{chiba}.  


\end{document}